%% file: aamas_main.tex
\documentclass[sigconf]{aamas}  

\usepackage{booktabs}
\usepackage{algorithm}
\usepackage{algpseudocode}
\usepackage[center]{subfigure}
\usepackage{epstopdf}
\usepackage{flushend}
\setcopyright{ifaamas}  
\acmDOI{doi}  
\acmISBN{}  
\acmConference[AAMAS'20]{Proc.\@ of the 19th International Conference on Autonomous Agents and Multiagent Systems (AAMAS 2020), B.~An, N.~Yorke-Smith, A.~El~Fallah~Seghrouchni, G.~Sukthankar (eds.)}{May 2020}{Auckland, New Zealand}  
\acmYear{2020}  
\copyrightyear{2020}  
\acmPrice{}  

\begin{document}

\title{Bird Flocking Inspired Control Strategy for Multi-UAV Collective Motion}  



%
\author{Xiyuan Liu}
\affiliation{%
  \institution{The Hong Kong University of Science and Technology}
}
\email{xliuba@connect.ust.hk}
\author{Li Qiu}
\affiliation{%
  \institution{The Hong Kong University of Science and Technology}
}
\email{eeqiu@ust.hk}
%
%
%
%
%
%

\begin{abstract}  
UAV collective motion has become a hot research topic in recent years. The realization of UAV collective motion, however, relied heavily on centralized control method and suffered from instability. Inspired by bird flocking theory, a control strategy for UAV collective motion with distributed measure and control methods was proposed in this study. In order to appropriately adjust the inter-agent distance suitable for realization, the control law based on bird flocking theory was optimized, and the convergence of velocities and collision avoidance properties were presented through simulation results. Furthermore, the stable collective motion of two UAVs using visual relative information only with proposed strategy in both indoor and outdoor GPS-denied environments were realized.
\end{abstract}

\keywords{Collective Motion; Flocking; UAV; Multi-Agent System}  

\maketitle


\input{aamas_body}


\bibliographystyle{ACM-Reference-Format}  
\bibliography{aamas_main}  

\end{document}

%% file: aamas_body.tex
\section{Introduction}

Researches on multi-UAV system have gradually gained attention in academia, due to their potential applications in remote transportation, surveillance \cite{inspection} and rescue missions \cite{earthquake}. Compared with a single and powerful agent, a group of agents are more fault tolerant and intelligent through communication and duty reallocation. Various control models for multi-UAV system have been proposed, such as leader-wingman \cite{Wingman}, artificial potential field \cite{Martin2014,Saber2006,olfati2002distributed}, virtual structure \cite{Virtual2008,Askari2015} and behavior based models \cite{Behavior2004,Vicsek2014}, etc. However, these models required the UAV group to maintain a fixed geometry when the entire system has reached a stable state. This makes altering formation shape inconvenient when the whole system encounters a dynamically changing environment. Though modified control functions have arisen, parameter optimization is always a difficult task.

Bird flocking inspired control models are the alternatives since their flexibility and robustness outperform those of formations \cite{lowcost}. In natural bird flocks, coordination is more emphasized than formation and is more beneficial for the colony. Each agent percepts local environment, interact with neighboring agents and makes decisions on its own. At present, constant velocity models have been discussed by many researchers \cite{Vicsek1995,Coordination2013}, however, they were not suitable for realization since the directions of agents were drastically changed. Though velocity alignment and cohesion problem have been resolved in \cite{CuckerSmale2007}, collision avoidance between agents could not be assured. In addition, the inter-agent distance was guaranteed to be greater than a lower bound \cite{CuckerDong2010}, never the less, the relative distance might get very large and was hard to tune. To this end, much emphasis has to be placed on the development of a control model preserving velocity alignment, regional cohesion and separation \cite{Reynolds1987,FixedTopology,Saber2004}.

From realization side, to achieve multi-UAV collective motion with centralized perception systems, such as motion capture or global positioning systems is a direct approach \cite{Swarm2018,Vicsek2018,MPC}. However, the increasing number of total agents will lead to a huge computation burden and communication delay for centralized processor, as well as limit their performance in unknown environments.

In this paper, the control law based on the bird flocking theory for multi-agent system with adjustable inter-agent distance is optimized, the onboard monocular camera is used as the only sensor for both perception and state estimation, and we have achieved two UAV's collective motion in both indoor and outdoor environments. The results provide a possible solution to implement a distributed multi-UAV system and demonstrate collective motion with decentralized sensors. At present, few research results have been addressed on this issue.

\section{Background}

Bird flocking or distributed behavior model was first studied in \cite{Reynolds1987} to animate the aggregate motion of birds in computer simulation. Three criteria, separation, cohesion and alignment were proposed to describe a stable flocking model. Alignment criterion requires each agent to match self’s velocity with neighboring flockmates in the mid range, separation criterion requires each agent to repel neighboring flockmates in the short range and cohesion criterion requires each agent to steer neighboring flockmates in the long range.

For the following second order system (\ref{eq:second}), authors in \cite{moreau2004stability,ren2005consensus,Coordination2013,CuckerSmale2007,CuckerDong2010} focused more on designing the interaction function $a_{ij}(x)$ to stabilize the inter-agent distances, while authors in \cite{Saber2006,olfati2002distributed,FixedTopology,tanner2007flocking} emphasized more on the gradient based schemes $V_i$. When the total number of agents was greater than ten, however, fragmentation might occur that a reference agent had to be introduced.

\begin{equation}\label{eq:second}
\begin{aligned}
  \dot{x}_i(t)&=\dot{v}_i(t)\\
  \dot{v}_i(t)&=\sum^k_{j=1}a_{ij}(x)(v_j-v_i)-\bigtriangledown x_i V_i
\end{aligned}
\end{equation}

First order system (\ref{eq:first}) was considered in \cite{Connectedness,Invariant,Stability}. With carefully designed attraction and repulsion functions, Lyapunov function candidates and LaSalle’s Invariant Principle, all the flock members were expected to converge to a constant
arrangement $(\dot{x}=0)$ within a hyper-ball. Separation was ensured in the first-order system, however, velocity alignment was less satisfied.

\begin{equation}\label{eq:first}
  \dot{x}_i(t)=\dot{u}_i(t)
\end{equation}

\cite{PNAS,KNN,CuckerDong2016} discovered that interaction between agents in flocking did not necessarily depend on the metric distance but rather on the topological distance. An average number of six to seven nearest neighbors were involved in the interaction instead of the whole neighbors within a fixed distance. In \cite{DynamicTopology}, it was shown that with modified potential function, where the potential was constant when the relative distance between two agents was larger than a certain limit, the stability of the flock was guaranteed if no agent was separated initially.

\section{Design of the Control Model}

\subsection{Proposition of the Control Law}

Considering a flock of $k$ agents whose behaviors are described by (\ref{eq:proposed_ui}, \ref{eq:proposed_af}) in continuous time with initial positions satisfying $d_0<\|x_i(0)-x_j(0)\|^2<d_1$ for all $i\neq j$. All agents update their states with the current information of the relative displacements and velocities between themselves and their neighboring agents. When $\beta\leq\frac{1}{2}$, the velocities of agents in the flock will converge to a common value without collision with others. The regulator $\Lambda(v)$ is designed to adjust the internal repulsion and attraction forces from neighboring agents. $\alpha, \beta, \theta, K, d_0, d_1$ are tunable parameters and $x_i, v_i$ represent the $i^{th}$ agent's position and velocity at time $t$ respectively.

\begin{equation}\label{eq:proposed_ui}
\begin{aligned}
\dot{x_i}(t)&=v_i(t)\\
\dot{v_i}(t)&=u_i(t)\\
u_i(t)&=\underbrace{\sum^k_{j=1}a_{ij}(x)(v_j-v_i)}_{\text{alignment term}}+\\
&\quad\underbrace{\Lambda(v)\sum_{j\neq i}f_0(\|x_i-x_j\|^2)(x_i-x_j)}_{\text{separation term}}+\\
&\quad\underbrace{\Lambda(v)\sum_{j\neq i}f_1(\|x_i-x_j\|^2)(x_j-x_i)}_{\text{cohesion term}}
\end{aligned}
\end{equation}

\begin{equation}\label{eq:proposed_af}
\begin{aligned}
a_{ij}(x)&=\frac{K}{(\sigma^2+\|x_i-x_j\|^2)^{\beta}}\\
\Lambda(v)&=(\frac{1}{k}\sum_{i>j}\|v_i-v_j\|^2)^{\frac{1}{2}}\\
f_0(x)&=\frac{1}{(x-d_0)^{\theta}}\\
f_1(x)&=\frac{1}{(x-d_1)^{\theta}}
\end{aligned}
\end{equation}

\subsection{Proof of the Collision Avoidance}
Here, the Lemmas and Propositions originated from~\cite{CuckerDong2010} are borrowed. For the detailed proof, interested readers are referred to the original paper.
\begin{proof}
Let $F_f$ be the $k$ by $k$ adjacency matrix with entries $f_{ij}=f_0(\|x_i-x_j\|^2)-f_1(\|x_i-x_j\|^2)$ when $i\neq j$ and $f_{ii}=0$. Let $D_f$ be a diagonal matrix whose entries are $d_{ii}=\sum_{j=1}^k f_{ij}$. We have Laplacian matrix $L_f=D_f-F_f$. Similarly we have $L_x=D_x-A_x$, where $a_{ij}$ is defined in (\ref{eq:proposed_af}). Then (\ref{eq:proposed_ui}) could be equally defined as
\begin{equation}\label{eq:continuous}
\begin{aligned}
\dot{\mathbf{x}}&=\mathbf{v}\\
\dot{\mathbf{v}}&=-L_x\mathbf{v}+\Lambda(v)L_f\mathbf{x}\\
\end{aligned}
\end{equation}

\noindent
where $\mathbf{x}=(x_1^T, x_2^T, \dots, x_k^T)^T$ and $\mathbf{v}=(v_1^T, v_2^T, \dots, v_k^T)^T$. Noticed that matrix $L_x$ acts on $\mathbb{E}^K$ by mapping $(v_1^T, v_2^T, \dots, v_k^T)^T$ to $((L_x)_{i1}v_1+\dots+(L_x)_{ik}v_k)_{i\leq k}$.

Let $\bigtriangleup=\{{(u^T, u^T, \dots, u^T)^T|u\in\mathbb{E}}\}$ be the diagonal of $\mathbb{E}^K$, and $\bigtriangleup^{\perp}$ be the orthogonal complement of $\bigtriangleup$ in $\mathbb{E}^K$. Then we could decompose every element $e\in\mathbb{E}^K=e_{\bigtriangleup}+e^{\perp}$ with $e_{\bigtriangleup}\in\bigtriangleup$ and $e^{\perp}\in\bigtriangleup^{\perp}$. We then decompose $\mathbf{v}=\mathbf{v_{\bigtriangleup}+v^{\perp}}$ where $\mathbf{v_{\bigtriangleup}}=(\bar{v}^T, \bar{v}^T, \dots, \bar{v}^T)^T$, $\mathbf{v^{\perp}}=(v_1^T-\bar{v}^T, \dots, v_k^T-\bar{v}^T)^T$ and $\bar{v}$ is the average velocity of all $k$ agents. We show that $\langle\mathbf{v_{\bigtriangleup}}, \mathbf{v^{\perp}}\rangle=\sum_{i=1}^k\langle\bar{v},v_i-\bar{v}\rangle=\langle\bar{v},\sum_{i=1}^k(v_i-\bar{v})\rangle=0$.

\begin{lemma}\label{lemma_1}
For any solution $(\mathbf{x}(t), \mathbf{v}(t))$ of (\ref{eq:continuous}), we have $\frac{d}{dt}\mathbf{v}_{\bigtriangleup}=0$.
\end{lemma}

\textsc{Lemma}~\ref{lemma_1} indicates that $\mathbf{v_{\bigtriangleup}}$ is constant with $t\geq0$ and $a_{ij}(x)=a_{ij}(x^{\perp})$. Thus in the following paragraphs, we use $\mathbf{x}$ and $\mathbf{v}$ to denote $\mathbf{x}^{\perp}$ and $\mathbf{v}^{\perp}$ respectively. The system in (\ref{eq:continuous}) then becomes the following

\begin{equation}\label{eq:continuous_projection}
\begin{aligned}
\dot{\mathbf{x}}&=\mathbf{v}\\
\dot{\mathbf{v}}&=-L_x\mathbf{v}+\|\mathbf{v}\|L_f\mathbf{x}.
\end{aligned}
\end{equation}

Define the function $E: \mathbb{E}^k\times\mathbb{E}^k\to(0,\infty)$ by (\ref{eq:Exv}), where $\delta$ is an infinitesimal positive number.

\begin{equation}\label{eq:Exv}
\begin{aligned}
E(\mathbf{x}, \mathbf{v})&=\|\mathbf{v}\|+\frac{1}{2}\sum_{i>j}\int_{\|x_i-x_j\|^2}^{d_1-\delta}f_0(r)dr-\frac{1}{2}\sum_{i>j}\int_{\|x_i-x_j\|^2}^{d_1-\delta}f_1(r)dr\\
&=\|\mathbf{v}\|+\frac{1}{2}\sum_{i>j}\int_{\|x_i-x_j\|^2}^{d_1-\delta}(f_0(r)-f_1(r))dr
\end{aligned}
\end{equation}

\begin{proposition}\label{prop1}
For all $t>0$, $-Hk\|\mathbf{v}\|+\langle L_f\mathbf{x}, \mathbf{v}\rangle\leq\frac{d}{dt}\|\mathbf{v}\|\leq-\frac{Hk\|\mathbf{v}\|}{(1+2\|\mathbf{x}\|^2)^{\beta}}+\langle L_f\mathbf{x}, \mathbf{v}\rangle$.
\end{proposition}

\begin{lemma}\label{lemma_2}
Let $A$ be a $k\times k$ positive, symmetric matrix and $L$ be its Laplacian. Then for all $u, w\in\mathbb{E}^K$ (and in particular, for all $u, w\in\bigtriangleup^{\perp}$), $\langle w, Lu\rangle=\sum_{i>j}\langle w_i-w_j, u_i-u_j\rangle a_{ij}$.
\end{lemma}

\noindent
Using \textsc{Proposition}~\ref{prop1}, \textsc{Lemma}~\ref{lemma_2} and the fundamental theorem of calculus we see the derivative of $E$ along the solution satisfies
\begin{equation}\label{eq:dExv}
\begin{aligned}
\frac{d}{dt}E(\mathbf{x}, \mathbf{v})&=\frac{d}{dt}\|\mathbf{v}\|+\frac{1}{2}\sum_{i>j}\frac{d}{dt}\int_{\|x_i-x_j\|^2}^{d_1-\delta}(f_0(r)-f_1(r))dr\\
&=\frac{d}{dt}\|\mathbf{v}\|-\frac{1}{2}\sum_{i>j}\frac{d}{dt}\langle x_i-x_j,x_i-x_j\rangle(f_0(\|x_i-x_j\|^2)\\
&\quad\ -f_1(\|x_i-x_j\|^2))\\
&\leq-\frac{Hk}{(1+2\|\mathbf{x}\|^2)^{\beta}}\|\mathbf{v}\|+\langle L_f\mathbf{x}, \mathbf{v}\rangle-\sum_{i>j}\langle x_i-x_j, v_i-v_j\rangle\\
&\quad\ (f_0(\|x_i-x_j\|^2)-f_1(\|x_i-x_j\|^2))\\
&=-\frac{Hk}{(1+2\|\mathbf{x}\|^2)^{\beta}}\|\mathbf{v}\|.
\end{aligned}
\end{equation}

\noindent
Hence $E(\mathbf{x}(t), \mathbf{v}(t))$ is a decreasing function of $t$ along the solution $(\mathbf{x}(t), \mathbf{v}(t))$. Write
\begin{equation}\label{eq:E0}
E(\mathbf{x}(0), \mathbf{v}(0))=\|\mathbf{v}(0)\|+\frac{1}{2}\sum_{i>j}\int_{\|x_i(0)-x_j(0)\|^2}^{d_1-\delta}(f_0(r)-f_1(r))dr.
\end{equation}

\noindent
Then, for all $t\geq0$, $E(\mathbf{x}(t), \mathbf{v}(t))\leq E(\mathbf{x}(0), \mathbf{v}(0))$. This implies
\begin{equation}
\frac{1}{2}\sum_{i>j}\int_{\|x_i-x_j\|^2}^{d_1-\delta}(f_0(r)-f_1(r))dr<E(\mathbf{x}(0), \mathbf{v}(0)).
\end{equation}

\noindent
From the definition of $f_0$ and $f_1$ and the initial conditions that we have ensured $d_0<\|x_i(t)-x_j(t)\|^2<d_1$ for all $i\neq j$ and $t\geq0$, otherwise the integral in (\ref{eq:Exv}) will go to infinity.
\end{proof}

\subsection{Proof of the Velocity Convergence}
\begin{proof}
From (\ref{eq:dExv}) we could obtain (\ref{eq:Et_E0}, \ref{eq:intE0})
\begin{equation}\label{eq:Et_E0}
E(\mathbf{x}(t), \mathbf{v}(t))-E(\mathbf{x}(0), \mathbf{v}(0))\leq-\int_0^t\frac{Hk}{(1+2\|\mathbf{x}(s)\|^2)^{\beta}}\|\mathbf{v}(s)\|ds
\end{equation}

\begin{equation}\label{eq:intE0}
\int_0^t\frac{Hk}{(1+2\|\mathbf{x}(s)\|^2)^{\beta}}\|\mathbf{v}(s)\|ds\leq E(\mathbf{x}(0), \mathbf{v}(0))
\end{equation}

\begin{proposition}\label{prop2}
For all $t>0$, $\frac{d}{dt}\|\mathbf{x}(t)\|\leq\|\mathbf{v}(t)\|$.
\end{proposition}

By \textsc{Proposition}~\ref{prop2}, we have
\begin{equation}\label{eq:proposition2}
\frac{d}{dt}\|\mathbf{x}(t)\|\leq\|\mathbf{v}(t)\|
\end{equation}

\begin{equation}
\int_{\|x(0)\|}^{\|x(t)\|}\frac{Hk}{(1+2y^2)^{\beta}}dy\leq E(\mathbf{x}(0), \mathbf{v}(0))
\end{equation}

\begin{equation}
\begin{aligned}
E(\mathbf{x}(0), \mathbf{v}(0))&\geq\int_{\|x(0)\|}^{\|x(t)\|}\frac{Hk}{(1+2y^2)^{\beta}}dy\\
&\geq\int_{\|x(0)\|}^{\|x(t)\|}\frac{Hky}{(1+2y^2)^{\beta+\frac{1}{2}}}dy\\
&=\left\{\begin{array}{rcl}\frac{Hk}{2-4\beta}(1+2y^2)^{\frac{1}{2}-\beta}|^{\|x(t)\|}_{\|x(0)\|}, & & \text{if $\beta\neq\frac{1}{2}$}\\\frac{Hk}{4}ln(1+2y^2)|^{\|x(t)\|}_{\|x(0)\|}, & & \text{if $\beta=\frac{1}{2}$}\end{array} \right.
\end{aligned}
\end{equation}

\noindent
Assume $\|\mathbf{x}(t)\|$ is unbounded for $t>0$, then as $t\to\infty$, $E(\mathbf{x}(0), \mathbf{v}(0))\to\infty$, which contradicts with our initial condition. Thus $\|\mathbf{x}(t)\|$ is bounded which leads to
\begin{equation}\label{eq:boundv}
\int_0^{\infty}\|\mathbf{v}(t)\|dt<\infty.
\end{equation}

\noindent
We show that $\|\mathbf{v}(t)\|$ is a continuous function of $t$ that
\begin{equation}
\begin{aligned}
|\frac{d}{dt}\|\mathbf{v}(t)\||&\leq Hk\|\mathbf{v}(t)\|+|\langle L_f\mathbf{x}(t), \mathbf{v}(t)\rangle|\\
&=Hk\|\mathbf{v}(t)\|\\
&\quad+|\sum_{i>j}f(\|x_i(t)-x_j(t)\|^2)\langle v_i-v_j, x_i-x_j\rangle|\\
&\leq Hk\|\mathbf{v}(t)\|\\
&\quad+\sum_{i>j}f(\|x_i(t)-x_j(t)\|^2)\|v_i-v_j\| \|x_i-x_j\|\leq\infty.
\end{aligned}
\end{equation}

\noindent
Then we show that when $t\to\infty$, $\|\mathbf{v}(t)\|\to0$. Assumes there exist some $\delta>0$ that when $t\to\infty$, $\|\mathbf{v}(t)\|\to\delta$, then
\begin{equation}
\int_0^{\infty}\|\mathbf{v}(t)\|dt\geq\infty
\end{equation}

\noindent
which contradicts (\ref{eq:boundv}).
\end{proof}

\section{Experiments}
\subsection{Simulation Test}

The effectiveness of the proposed control law is demonstrated in two scenarios, leaderless flocking and flocking with leader. In the first case, all agents are treated equally and for each agent, the relative information between itself and others are used. In the second case, only leader agent is moving randomly regardless of neighboring agents, and the follower agents are still manipulated by the control law. The parameters of (\ref{eq:proposed_ui}, \ref{eq:proposed_af}) used in simulation are summarized in Table~\ref{tab:initial_param}.

\begin{table}[htb]
  \caption{Parameters in Simulation Test}
  \label{tab:initial_param}
  \begin{tabular}{ccccccc}
    \toprule
    Parameter&$\sigma$&$\beta$&$\theta$&$K$&$d_0$&$d_1$\\
    \midrule
    Value&1&0.5&2&1&1&2.25\\
  \bottomrule
  \end{tabular}
\end{table}

\subsubsection{Flocking without Leader}

\begin{figure*}
\centering
\subfigure[Trajectories of leaderless flocking.]{\label{fig:N3_pos}\includegraphics[width=0.33\textwidth]{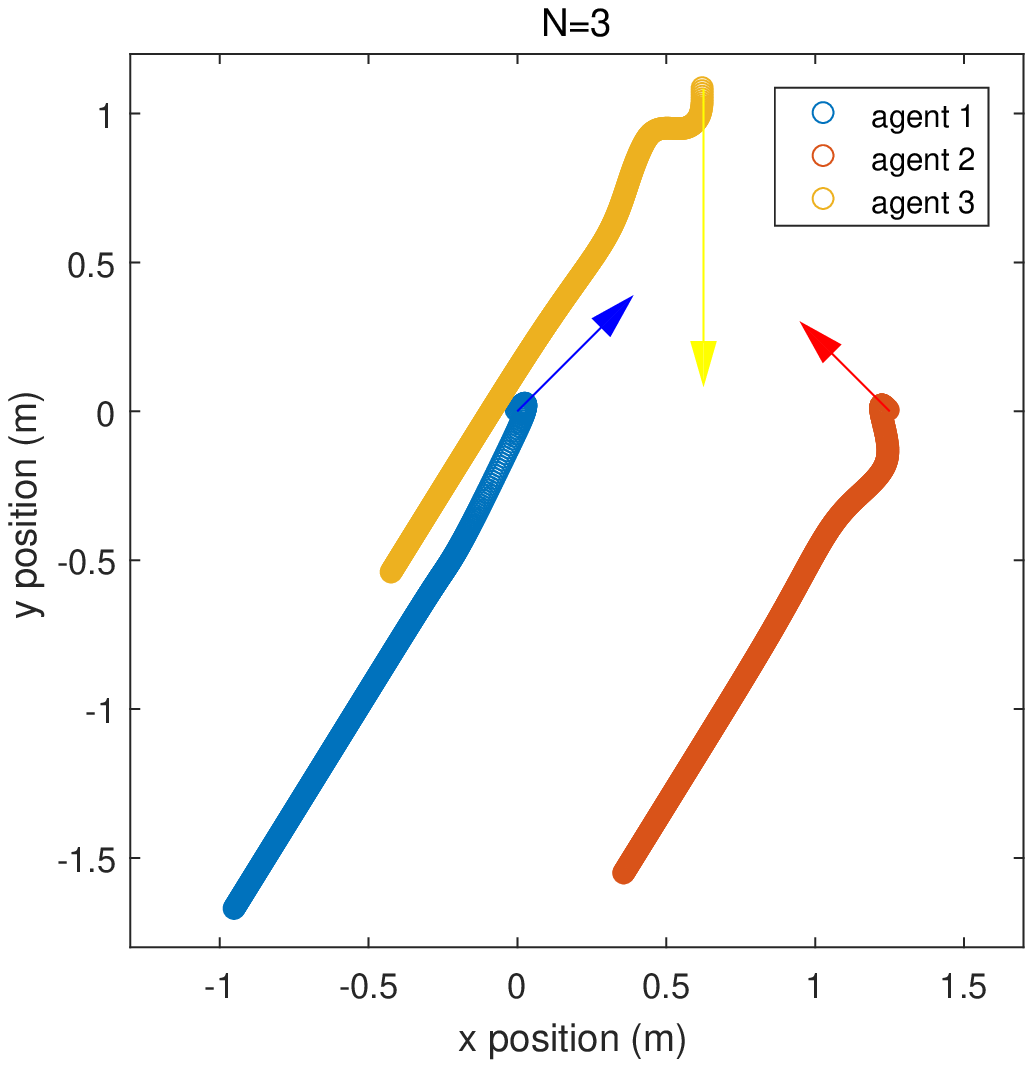}}
\subfigure[Magnitude of velocities of leaderless flocking.]{\label{fig:N3_mag}\includegraphics[width=0.33\textwidth]{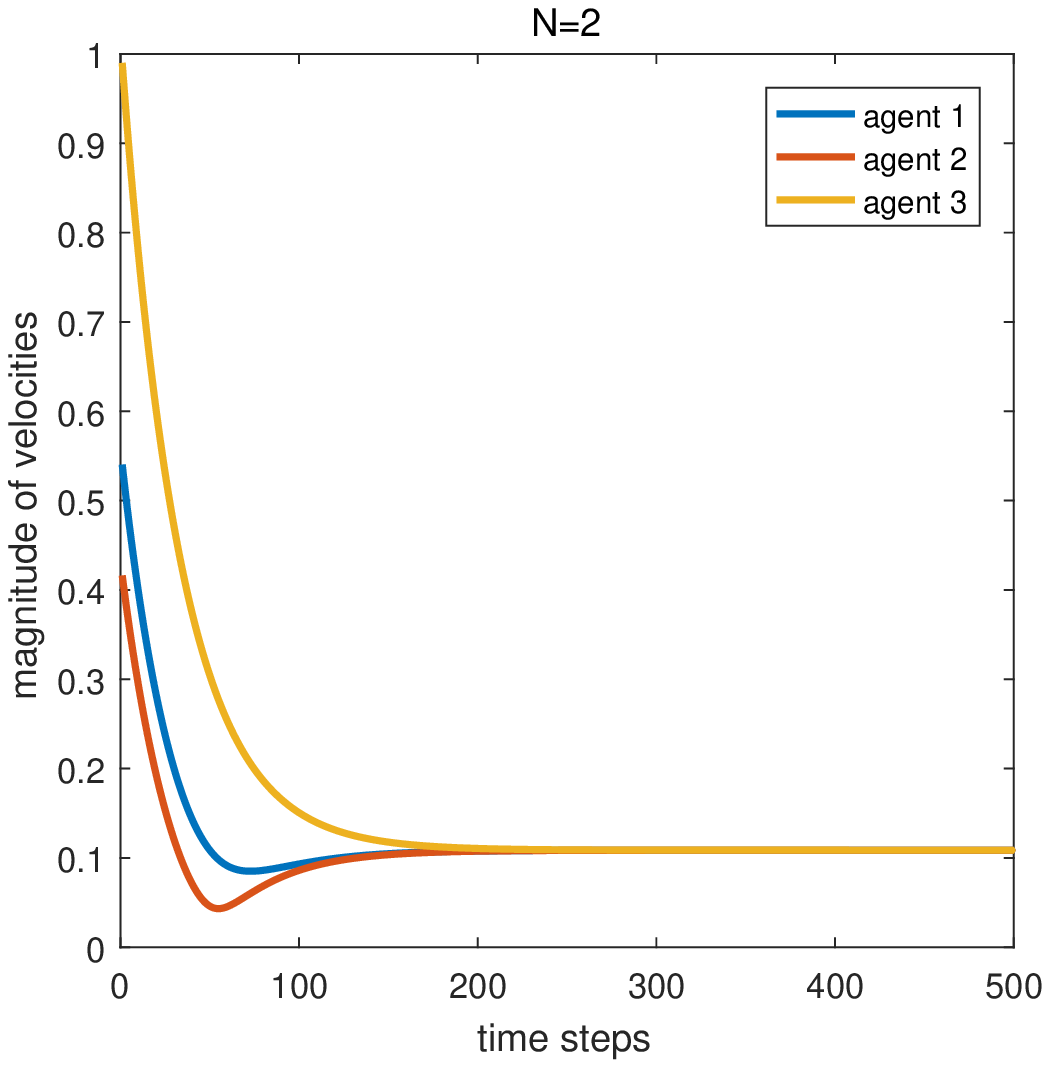}}
\subfigure[Inter-agent distances of leaderless flocking.]{\label{fig:N3_dis}\includegraphics[width=0.33\textwidth]{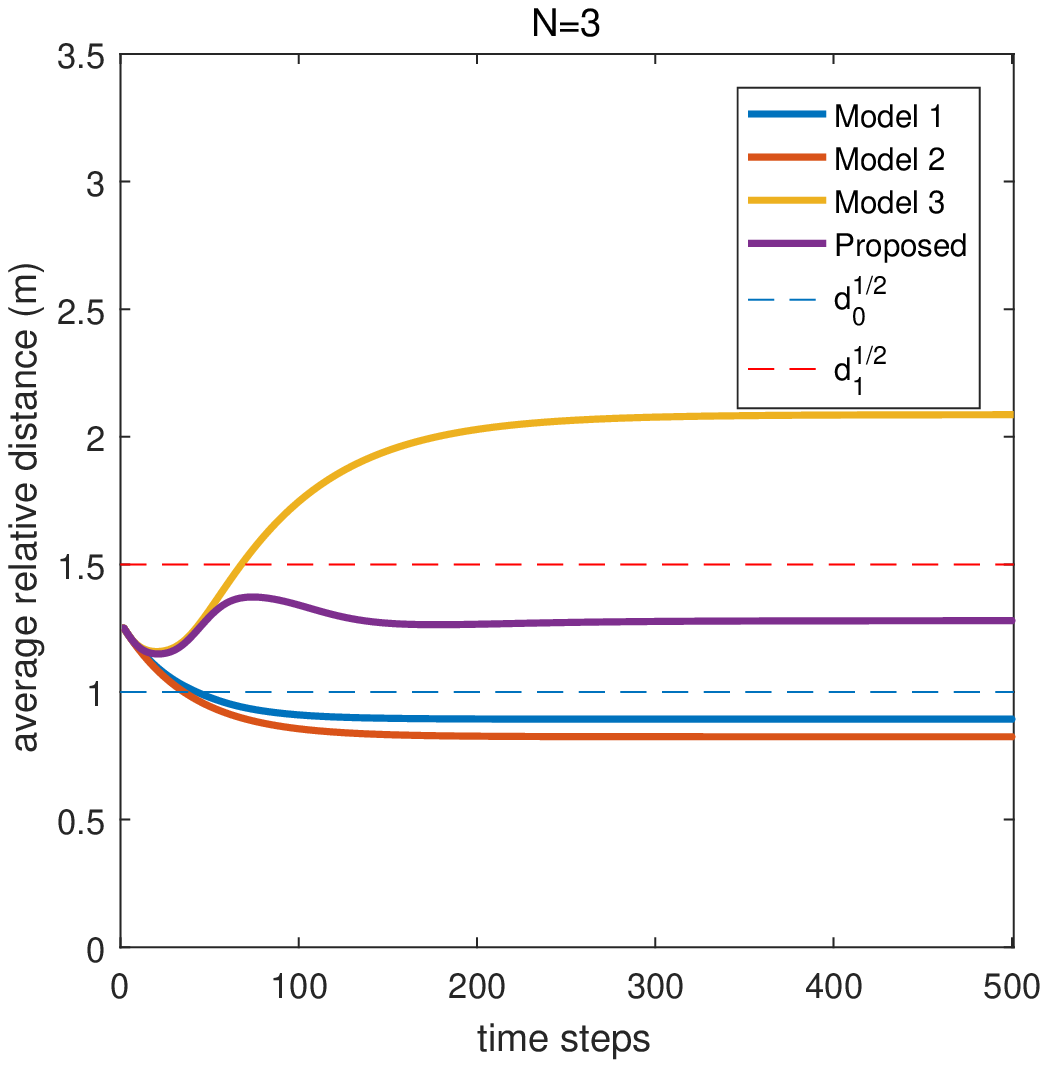}}
\caption{Leaderless flocking of three agents. \ref{fig:N3_pos} The movement direction of each agent quickly converges to a common value. \ref{fig:N3_mag} The magnitude of each agent's velocity quickly converges to a common value. \ref{fig:N3_dis} The average inter-agent distance of model 3 is beyond the upper bound, and the those of model 1 and 2 are beneath the lower bound.}\label{fig:N3}
\end{figure*}

Consider a group of three agents with initial conditions listed in Table~\ref{tab:simu1}. Their initial orientations are specially designed to face to their geometrical center, to demonstrate the collision avoidance property and the inter-agent distance adjustment capability, as shown in Fig.~\ref{fig:N3_pos}. The length of the colored arrow indicates the magnitude of each agent's initial velocity. Their velocities and average inter-agent distances are illustrated in Fig.~\ref{fig:N3_mag} and Fig.~\ref{fig:N3_dis}. It is shown that our control law properly maintain the inter-agent distance within the boundaries, compared with model 1 in \cite{Vicsek1995}, model 2 in \cite{CuckerSmale2007} and model 3 in \cite{CuckerDong2010}. This property is crucial for the platform setup and realization in narrow space, since the relative distance should not beyond the camera's maximum operating range, and also should not beneath UAV's minimum safety range.

\begin{table}[htb]
  \caption{Initial Parameters for Flocking without Leader Simulation}
  \label{tab:simu1}
  \begin{tabular}{cccc}
    \toprule
    Number&Position(m)&Orientation(deg)&Velocity(m/s)\\
    \midrule
    Agent 1 & (0, 0) & 45 & 0.54\\
    Agent 2 & (1.25, 0) & 135 & 0.42\\
    Agent 3 & (0.63, 1.08) & 270 & 0.99\\
  \bottomrule
  \end{tabular}
\end{table}

\subsubsection{Flocking with Leader}

To cooperate with the experiment setup, only one leader and one follower scenario is considered. The leader UAV is driven by the following desired input:

\begin{displaymath}
  u_x(t)=\left\{
\begin{array}{rcl}
sin(\frac{\pi t}{180}) & & {t < 125}\\
-sin(\frac{\pi t}{180}) & & {125 \leq t}
\end{array} \right.
\end{displaymath}

\begin{displaymath}
  u_y(t)=\left\{
\begin{array}{rcl}
cos(\frac{\pi t}{180}) & & {t < 125}\\
-cos(\frac{\pi t}{180}) & & {125 \leq t}
\end{array} \right.
\end{displaymath}

\begin{figure*}
\centering
\subfigure[Trajectories of leader-follower flocking.]{\label{fig:N2_pos}\includegraphics[width=0.33\textwidth]{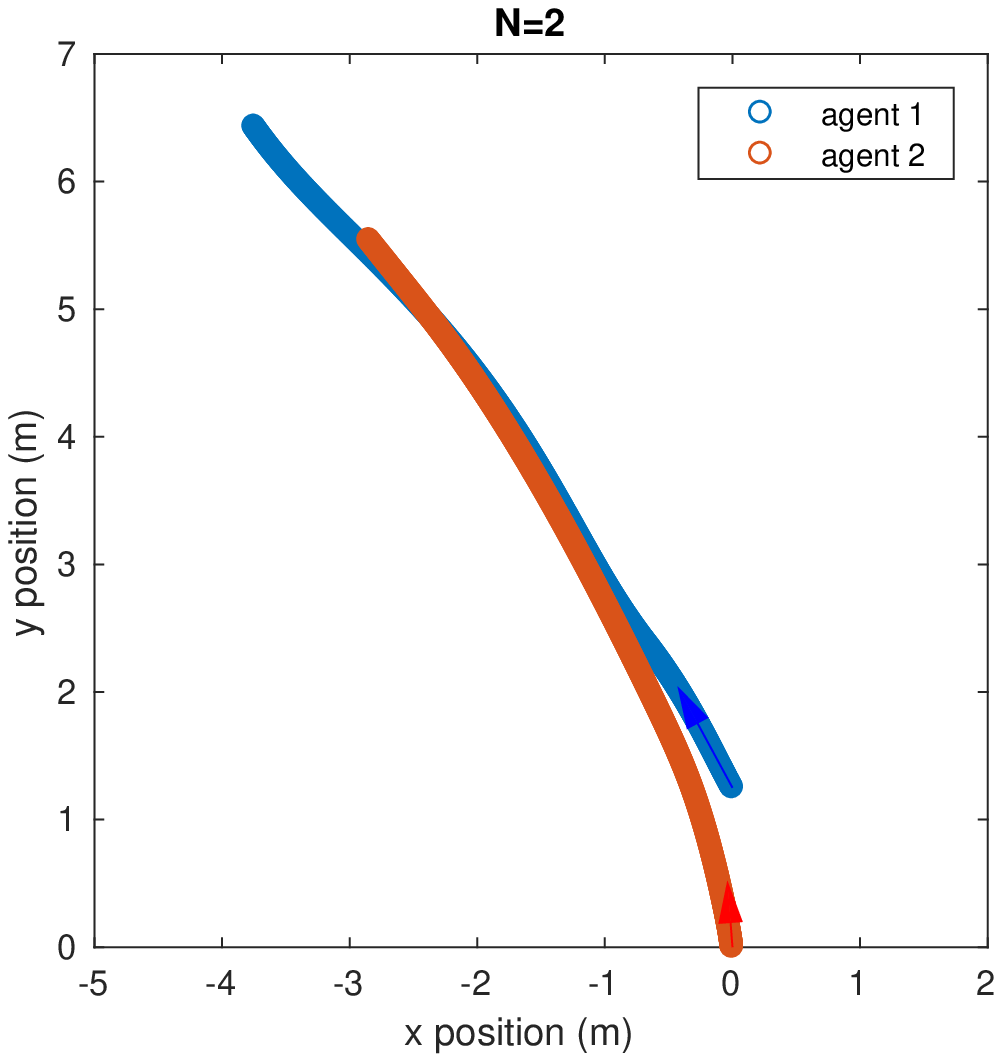}}
\subfigure[Magnitude of velocities of leader-follower flocking.]{\label{fig:N2_mag}\includegraphics[width=0.33\textwidth]{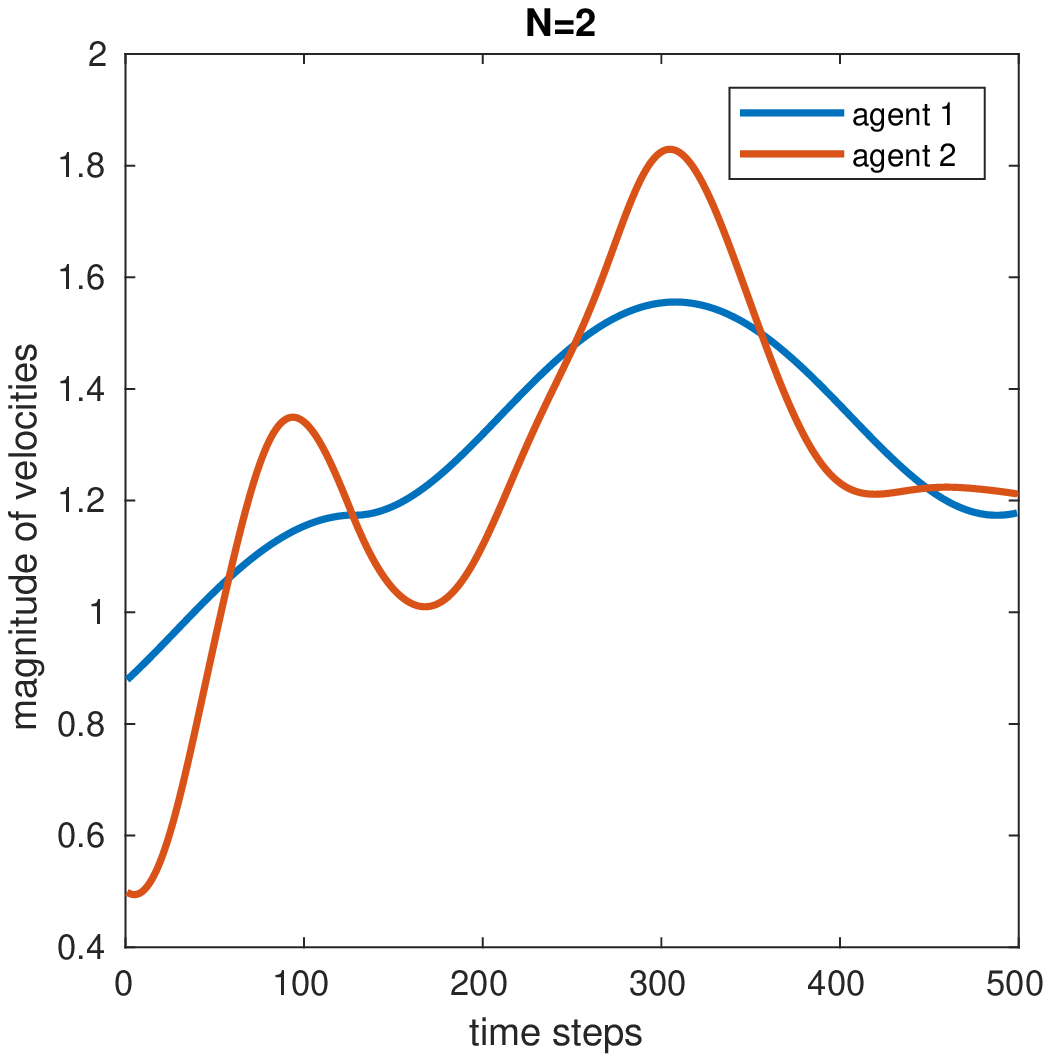}}
\subfigure[Inter-agent distances of leader-follower flocking.]{\label{fig:N2_dis}\includegraphics[width=0.33\textwidth]{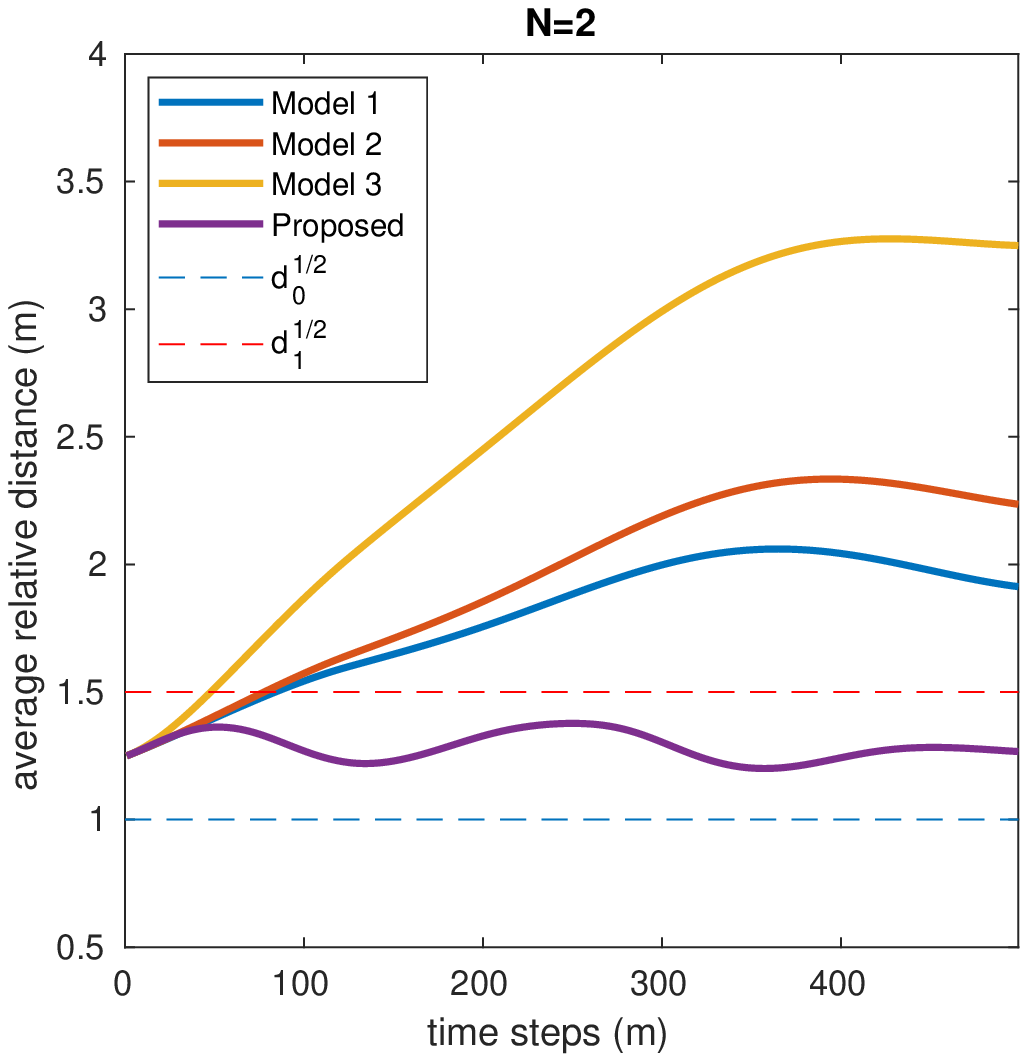}}
\caption{One leader one follower flocking. \ref{fig:N2_pos} The follower agent 2 smoothly tracks the leader agent 1 with the control law. \ref{fig:N2_mag} The follower agent 2's velocity is properly manipulated to stay flocking with agent 1. \ref{fig:N2_dis} Agents controlled by model 1, 2 and 3 fail to track the leader agent within an appropriate range, and the proposed control law handle the inter-agent distance suitably.}\label{fig:N2}
\end{figure*}

The initial condition of two agents is listed in Table \ref{tab:simu2}. The results show that the movement of two agents are still in coordination and the inter-agent distance is carefully bounded as illustrated in Fig.~\ref{fig:N2}.

\begin{table}[htb]
  \caption{Initial Parameters for Leader-Follower Flocking Simulation}
  \label{tab:simu2}
  \begin{tabular}{cccc}
    \toprule
    Number&Position(m)&Orientation(deg)&Velocity(m/s)\\
    \midrule
    Agent 1 & (0, 1.32) & 118 & 0.88\\
    Agent 2 & (0, 0) & 92 & 0.50\\
  \bottomrule
  \end{tabular}
\end{table}

\subsection{UAV Platform Construction}

The quadrotor is chosen as the platform due to its light weight and agility in maneuvering in confined space to realize the multi-UAV collective motion. Unlike many existed flocking systems that rely on motion capture system (VICON), global positioning system (RTK or GPS), in this study, the onboard camera is used as the only sensor both to imitate natural birds and exhibit the decentralized measure method. The results through both indoor and outdoor experiments show that the minimal sensing setup is sufficient for the realization of proposed control law.

\subsubsection{Hardware Implementation}

\begin{figure}
\includegraphics[width=0.47\textwidth]{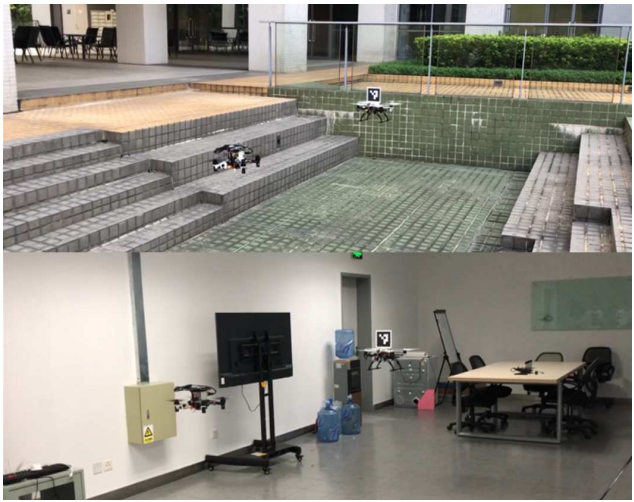}
\caption{UAV flocking system in indoor (top) and outdoor (bottom) environments. The front one (with tag) is the leader and the back one is the follower.}
\label{fig:outdoor}
\end{figure}

Two quadrotors are constructed with one being leader and another one being follower as shown in Fig.~\ref{fig:outdoor}. The front leader UAV is constructed on F330 frame and carries DJI Guidance system.\footnote{https://www.dji.com/hk-en/guidance} The follower UAV is constructed on Q250 frame and carries Intel RealSense D435i camera,\footnote{https://www.intelrealsense.com/depth-camera-d435i/} Intel NUC i5 Mini-Computer.\footnote{https://www.intel.com/content/www/us/en/products/boards-kits/nuc/boards/nuc7i5dnbe.html} Both UAVs are armed with DJI N3 flight controller for low level control.\footnote{https://www.dji.com/hk-en/n3} The total weight of the leader and follower UAV are 1.08 kg and 1.12 kg respectively. More details about the platform used in this study could be found in Table~\ref{tab:hardware_power}.

\begin{table}
  \caption{Weight and Power Consumption of Components on the Follower UAV}
  \label{tab:hardware_power}
  \begin{tabular}{ccc}
    \toprule
    Components&Weight(g)&Power(W)\\
    \midrule
    Intel NUC i5 mini computer & 125 & 15(avg)\\
    Intel Realsense D435i camera & 272 & 2.5(avg)\\
    DJI N3 flight controller & 92 & 3.3(avg)/5(peak)\\
    DJI Lightbridge 2 receiver & 70 & 7.8(avg)\\
  \bottomrule
  \end{tabular}
\end{table}

Two UAVs are constructed in this study mainly due to the initial goal was to fill in the gap of the realization of flocking model on UAVs with distributed control algorithm and measurement method, and the field-of-view of one forward looking camera has limited the total number of UAVs. To extend the current setup to multiple UAVs, two fish-eye cameras (pointing forward-and-backward or left-and-right) or one omnidirectional camera (looking upward) could be used to cover a wider FOV, such that no blind region will exist in each agent' perception system. We leave this for our future work.

\subsubsection{Software Implementation}

\begin{figure}
\includegraphics[width=0.47\textwidth]{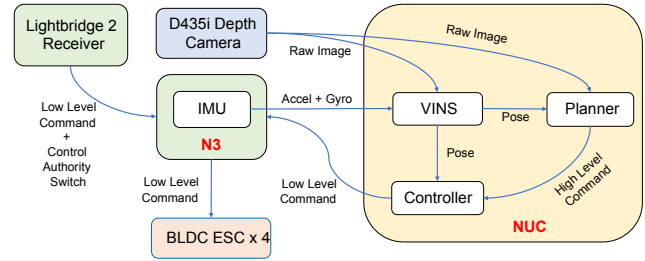}
\caption{Software architecture on follower UAV.}
\label{fig:software_architecture}
\end{figure}

The software architecture on follower UAV is shown in Fig.~\ref{fig:software_architecture}. We implement our algorithm with C++ programming language under ROS environment.\footnote{www.ros.org} On the follower's mini i5 computer, 400 Hz IMU measurements and 30 Hz grey scale image data are fused in the visual-inertial state estimator~\cite{VINS} to obtain UAV self position and orientation. Unique aruco code~\cite{Aruco} is attached on leader UAV to simplify the relative displacement and pose estimation. The high level flocking algorithm is running at 10 Hz in the planner, as shown in Algorithm~\ref{alg:alg_flock}. The low level command is handled by flight controller and published at 400 Hz.

\begin{algorithm}[h]
  \caption{Flocking Algorithm for Each Agent UAV}
  \label{alg:alg_flock}
  \begin{algorithmic}[1]
    \If{flocking signal triggered}
      \State obtain body pose estimation from VINS;
      \State capture leader's image;
      \While {new image}
        \State estimate relative displacement $\mathit{x_i-x_j}$;
        \State estimate relative velocity $\mathit{v_i-v_j}$;
        \State calculate desired $\mathit{u_i}$ from proposed model (\ref{eq:proposed_ui}, \ref{eq:proposed_af});
        \State execute flocking command;
      \EndWhile
    \EndIf
  \end{algorithmic}
\end{algorithm}

\subsection{Real World Experiment}

\begin{figure}[htb]
\includegraphics[width=0.40\textwidth]{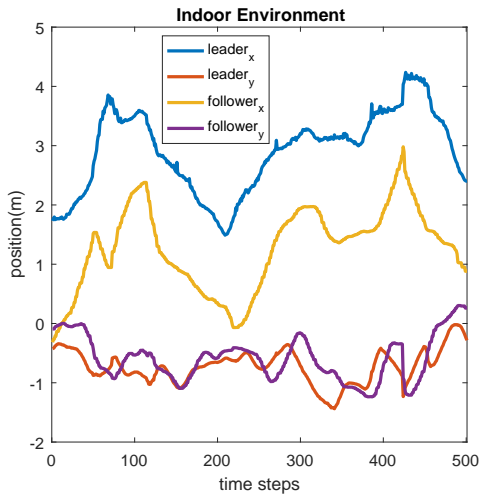}
\caption{Position profile in indoor environment.}
\label{fig:indoor_xy}
\end{figure}

\begin{figure}[htb]
\includegraphics[width=0.40\textwidth]{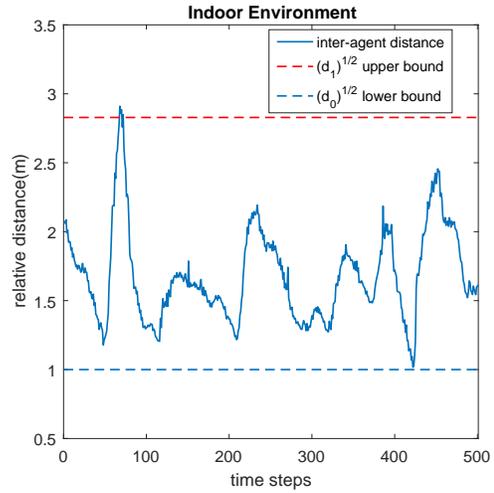}
\caption{Relative distance profile in indoor environment.}
\label{fig:indoor_dis}
\end{figure}

\begin{figure}[htb]
\includegraphics[width=0.40\textwidth]{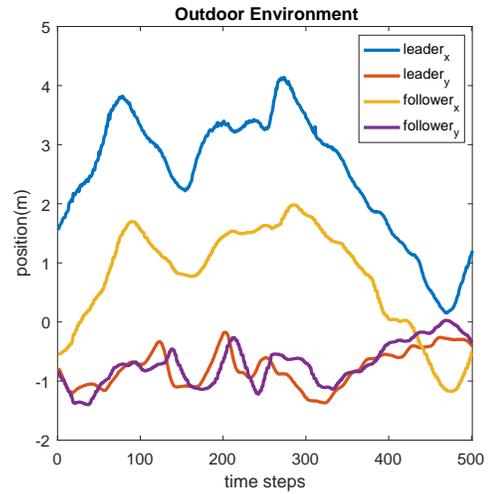}
\caption{Position profile in outdoor environment.}
\label{fig:outdoor_xy}
\end{figure}

\begin{figure}[htb]
\includegraphics[width=0.40\textwidth]{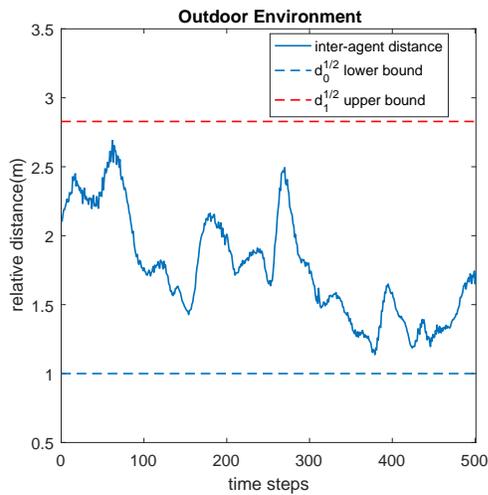}
\caption{Relative distance profile in outdoor environment.}
\label{fig:outdoor_dis}
\end{figure}

In this section, we demonstrate that our flocking system is able to operate in both indoor and outdoor GPS-denied environments. The desired $u_i$ is updated when a newer photo comes in and in each time interval (roughly 0.1 s) the latest $u_i$ is being executed. The initial take off position of follower UAV is 1.5 m behind the leader UAV to match the requirement ($d_0<||x_i-x_j||^2<d_1$) with parameters $d_0=1, d_1=8, K=1, k=2, \alpha=1$ and $\beta=0.25$ in (\ref{eq:proposed_ui}, \ref{eq:proposed_af}). The maximum acceleration and velocity of follower UAV are set to $a_{max}=2.5 m/s^2$ and $v_{max}=0.5 m/s$ for safety reasons. In each experiment, the position and average relative distance profiles are illustrated in Fig.~\ref{fig:indoor_xy}, \ref{fig:indoor_dis}, \ref{fig:outdoor_xy}, \ref{fig:outdoor_dis} with data captured from VINS. Due to the measurement noise, light and wind condition, spikes appear randomly in all figures. We have shown that the trend of position in x-y direction of follower UAV is in accordance with that of leader, the follower UAV is able stay flocking with the leader and the reaction of follower is rapid.

\section{Conclusion and Future Work}

In this thesis, a flocking system consisting two UAVs with the control law based on bird flocking theory is presented and the detailed hardware and software architectures are introduced. The monocular camera is used as the only on-board sensor for both follower UAV’s state estimation and leader UAV’s pose recognition. When in flocking, the follower UAV first percepts the surrounding environment to estimate self state, then recognizes leader UAV’s pose and calculates desired acceleration using this proposed model and third executes the desired input until next image is captured and processed.

Simulations and real world experiments have been conducted and analysed. The results show that this flocking system has met the three flocking criteria without relying on any external perception system or centralized control panel, achieved fast convergence rate and kept bounded relative distance with neighboring agents to avoid collision, and also given fine weather conditions.

Our future work will focus on the flocking of more than two UAVs in GPS-denied environment, including extending this proposed flocking model from fixed topology to dynamic topology, extending the flocking model form homogeneous to heterogeneous, introducing multiple fisheye cameras or an omnidirectional camera for perception and implementing ultra wide band (UWB) sensor for relative displacement measurement and internal communication.

%% file: aamas_main.bbl

\begin{thebibliography}{00}


\ifx \showCODEN    \undefined \def \showCODEN     #1{\unskip}     \fi
\ifx \showDOI      \undefined \def \showDOI       #1{#1}\fi
\ifx \showISBNx    \undefined \def \showISBNx     #1{\unskip}     \fi
\ifx \showISBNxiii \undefined \def \showISBNxiii  #1{\unskip}     \fi
\ifx \showISSN     \undefined \def \showISSN      #1{\unskip}     \fi
\ifx \showLCCN     \undefined \def \showLCCN      #1{\unskip}     \fi
\ifx \shownote     \undefined \def \shownote      #1{#1}          \fi
\ifx \showarticletitle \undefined \def \showarticletitle #1{#1}   \fi
\ifx \showURL      \undefined \def \showURL       {\relax}        \fi
\providecommand\bibfield[2]{#2}
\providecommand\bibinfo[2]{#2}
\providecommand\natexlab[1]{#1}
\providecommand\showeprint[2][]{arXiv:#2}

\bibitem[\protect\citeauthoryear{Askari, Mortazavi, and Talebi}{Askari
  et~al\mbox{.}}{2013}]%
        {Askari2015}
\bibfield{author}{\bibinfo{person}{A. Askari}, \bibinfo{person}{M. Mortazavi},
  {and} \bibinfo{person}{H.~A. Talebi}.} \bibinfo{year}{2013}\natexlab{}.
\newblock \showarticletitle{UAV formation control via the virtual structure
  approach}.
\newblock \bibinfo{journal}{{\em Journal of Aerospace Engineering\/}}
  \bibinfo{volume}{28}, \bibinfo{number}{1} (\bibinfo{year}{2013}),
  \bibinfo{pages}{04014047}.
\newblock


\bibitem[\protect\citeauthoryear{Baca, Loianno, and Saska}{Baca
  et~al\mbox{.}}{2016}]%
        {MPC}
\bibfield{author}{\bibinfo{person}{T. Baca}, \bibinfo{person}{G. Loianno},
  {and} \bibinfo{person}{M. Saska}.} \bibinfo{year}{2016}\natexlab{}.
\newblock \showarticletitle{Embedded model predictive control of unmanned micro
  aerial vehicles}. In \bibinfo{booktitle}{{\em International Conference on
  Methods and Models in Automation and Robotics (MMAR)}}. IEEE,
  \bibinfo{pages}{992--997}.
\newblock


\bibitem[\protect\citeauthoryear{Ballerini, Cabibbo, Candelier, Cavagna,
  Cisbani, Giardina, Lecomte, Orlandi, Parisi, and Procaccini}{Ballerini
  et~al\mbox{.}}{2008}]%
        {PNAS}
\bibfield{author}{\bibinfo{person}{M. Ballerini}, \bibinfo{person}{N. Cabibbo},
  \bibinfo{person}{R. Candelier}, \bibinfo{person}{A. Cavagna},
  \bibinfo{person}{E. Cisbani}, \bibinfo{person}{I. Giardina},
  \bibinfo{person}{V. Lecomte}, \bibinfo{person}{A. Orlandi},
  \bibinfo{person}{G. Parisi}, {and} \bibinfo{person}{A. Procaccini}.}
  \bibinfo{year}{2008}\natexlab{}.
\newblock \showarticletitle{Interaction ruling animal collective behavior
  depends on topological rather than metric distance: Evidence from a field
  study}.
\newblock \bibinfo{journal}{{\em Proceedings of the national academy of
  sciences\/}} \bibinfo{volume}{105}, \bibinfo{number}{4}
  (\bibinfo{year}{2008}), \bibinfo{pages}{1232--1237}.
\newblock


\bibitem[\protect\citeauthoryear{Chung, Paranjape, Dames, Shen, and
  Kumar}{Chung et~al\mbox{.}}{2018}]%
        {Swarm2018}
\bibfield{author}{\bibinfo{person}{S. Chung}, \bibinfo{person}{A.~A.
  Paranjape}, \bibinfo{person}{P. Dames}, \bibinfo{person}{S. Shen}, {and}
  \bibinfo{person}{V. Kumar}.} \bibinfo{year}{2018}\natexlab{}.
\newblock \showarticletitle{A survey on aerial swarm robotics}.
\newblock \bibinfo{journal}{{\em IEEE Transactions on Robotics\/}}
  \bibinfo{volume}{34}, \bibinfo{number}{4} (\bibinfo{year}{2018}),
  \bibinfo{pages}{837--855}.
\newblock


\bibitem[\protect\citeauthoryear{Cucker and Dong}{Cucker and Dong}{2010}]%
        {CuckerDong2010}
\bibfield{author}{\bibinfo{person}{F. Cucker} {and} \bibinfo{person}{J. Dong}.}
  \bibinfo{year}{2010}\natexlab{}.
\newblock \showarticletitle{Avoiding collisions in flocks}.
\newblock \bibinfo{journal}{{\it IEEE Trans. Automat. Control}}
  \bibinfo{volume}{55}, \bibinfo{number}{5} (\bibinfo{year}{2010}),
  \bibinfo{pages}{1238--1243}.
\newblock


\bibitem[\protect\citeauthoryear{Cucker and Dong}{Cucker and Dong}{2016}]%
        {CuckerDong2016}
\bibfield{author}{\bibinfo{person}{F. Cucker} {and} \bibinfo{person}{J. Dong}.}
  \bibinfo{year}{2016}\natexlab{}.
\newblock \showarticletitle{On flocks influenced by closest neighbors}.
\newblock \bibinfo{journal}{{\em Mathematical Models and Methods in Applied
  Sciences\/}} \bibinfo{volume}{26}, \bibinfo{number}{14}
  (\bibinfo{year}{2016}), \bibinfo{pages}{2685--2708}.
\newblock


\bibitem[\protect\citeauthoryear{Cucker and Smale}{Cucker and Smale}{2007}]%
        {CuckerSmale2007}
\bibfield{author}{\bibinfo{person}{F. Cucker} {and} \bibinfo{person}{S.
  Smale}.} \bibinfo{year}{2007}\natexlab{}.
\newblock \showarticletitle{Emergent behavior in flocks}.
\newblock \bibinfo{journal}{{\em IEEE Transactions on automatic control\/}}
  \bibinfo{volume}{52}, \bibinfo{number}{5} (\bibinfo{year}{2007}),
  \bibinfo{pages}{852--862}.
\newblock


\bibitem[\protect\citeauthoryear{Dehghani and Menhaj}{Dehghani and
  Menhaj}{2016}]%
        {Wingman}
\bibfield{author}{\bibinfo{person}{M.~A. Dehghani} {and} \bibinfo{person}{M.~B.
  Menhaj}.} \bibinfo{year}{2016}\natexlab{}.
\newblock \showarticletitle{Communication free leader–follower formation
  control of unmanned aircraft systems}.
\newblock \bibinfo{journal}{{\em Robotics and Autonomous Systems\/}}
  \bibinfo{volume}{80} (\bibinfo{year}{2016}), \bibinfo{pages}{69–75}.
\newblock
\showDOI{%
\url{https://doi.org/10.1016/j.robot.2016.03.008}}


\bibitem[\protect\citeauthoryear{Dimarogonas and Kyriakopoulos}{Dimarogonas and
  Kyriakopoulos}{2008}]%
        {Connectedness}
\bibfield{author}{\bibinfo{person}{D.~V. Dimarogonas} {and}
  \bibinfo{person}{K.~J. Kyriakopoulos}.} \bibinfo{year}{2008}\natexlab{}.
\newblock \showarticletitle{Connectedness preserving distributed swarm
  aggregation for multiple kinematic robots}.
\newblock \bibinfo{journal}{{\em IEEE Transactions on Robotics\/}}
  \bibinfo{volume}{24}, \bibinfo{number}{5} (\bibinfo{year}{2008}),
  \bibinfo{pages}{1213--1223}.
\newblock


\bibitem[\protect\citeauthoryear{Dimarogonas and Kyriakopoulos}{Dimarogonas and
  Kyriakopoulos}{2009}]%
        {Invariant}
\bibfield{author}{\bibinfo{person}{D.~V. Dimarogonas} {and}
  \bibinfo{person}{K.~J. Kyriakopoulos}.} \bibinfo{year}{2009}\natexlab{}.
\newblock \showarticletitle{Inverse agreement protocols with application to
  distributed multi-agent dispersion}.
\newblock \bibinfo{journal}{{\it IEEE Trans. Automat. Control}}
  \bibinfo{volume}{54}, \bibinfo{number}{3} (\bibinfo{year}{2009}),
  \bibinfo{pages}{657--663}.
\newblock


\bibitem[\protect\citeauthoryear{Dong and Sun}{Dong and Sun}{2004}]%
        {Behavior2004}
\bibfield{author}{\bibinfo{person}{M. Dong} {and} \bibinfo{person}{Z. Sun}.}
  \bibinfo{year}{2004}\natexlab{}.
\newblock \showarticletitle{A behavior-based architecture for unmanned aerial
  vehicles}. In \bibinfo{booktitle}{{\em Proceedings of the 2004 IEEE
  International Symposium on Intelligent Control}}. IEEE,
  \bibinfo{pages}{149--155}.
\newblock


\bibitem[\protect\citeauthoryear{Gazi and Passino}{Gazi and Passino}{2002}]%
        {Stability}
\bibfield{author}{\bibinfo{person}{V. Gazi} {and} \bibinfo{person}{K.~M.
  Passino}.} \bibinfo{year}{2002}\natexlab{}.
\newblock \showarticletitle{Stability analysis of swarms}. In
  \bibinfo{booktitle}{{\em Proceedings of the 2002 American Control
  Conference}}, Vol.~\bibinfo{volume}{3}. IEEE, \bibinfo{pages}{1813--1818}.
\newblock


\bibitem[\protect\citeauthoryear{Jadbabaie, Lin, and Morse}{Jadbabaie
  et~al\mbox{.}}{2003}]%
        {Coordination2013}
\bibfield{author}{\bibinfo{person}{A. Jadbabaie}, \bibinfo{person}{J. Lin},
  {and} \bibinfo{person}{A.~S. Morse}.} \bibinfo{year}{2003}\natexlab{}.
\newblock \showarticletitle{Coordination of groups of mobile autonomous agents
  using nearest neighbor rules}.
\newblock \bibinfo{journal}{{\em Departmental Papers (ESE)\/}}
  (\bibinfo{year}{2003}), \bibinfo{pages}{29}.
\newblock


\bibitem[\protect\citeauthoryear{Martin}{Martin}{2014}]%
        {KNN}
\bibfield{author}{\bibinfo{person}{S. Martin}.}
  \bibinfo{year}{2014}\natexlab{}.
\newblock \showarticletitle{Multi-agent flocking under topological
  interactions}.
\newblock \bibinfo{journal}{{\em Systems \& Control Letters\/}}
  \bibinfo{volume}{69} (\bibinfo{year}{2014}), \bibinfo{pages}{53--61}.
\newblock


\bibitem[\protect\citeauthoryear{Michael, Shen, Mohta, Kumar, and
  et~al.}{Michael et~al\mbox{.}}{2013}]%
        {earthquake}
\bibfield{author}{\bibinfo{person}{N. Michael}, \bibinfo{person}{S. Shen},
  \bibinfo{person}{K. Mohta}, \bibinfo{person}{V. Kumar}, {and}
  \bibinfo{person}{et al.}} \bibinfo{year}{2013}\natexlab{}.
\newblock \showarticletitle{Collaborative mapping of an earthquake damaged
  building via ground and aerial robots}.
\newblock \bibinfo{journal}{{\em Springer Tracts in Advanced Robotics Field and
  Service Robotics\/}} (\bibinfo{year}{2013}), \bibinfo{pages}{33–47}.
\newblock
\showDOI{%
\url{https://doi.org/10.1007/978-3-642-40686-7_3}}


\bibitem[\protect\citeauthoryear{Moreau}{Moreau}{2004}]%
        {moreau2004stability}
\bibfield{author}{\bibinfo{person}{L. Moreau}.}
  \bibinfo{year}{2004}\natexlab{}.
\newblock \showarticletitle{Stability of continuous-time distributed consensus
  algorithms}. In \bibinfo{booktitle}{{\em IEEE conference on decision and
  control (CDC)}}, Vol.~\bibinfo{volume}{4}. IEEE, \bibinfo{pages}{3998--4003}.
\newblock


\bibitem[\protect\citeauthoryear{Olfati-Saber}{Olfati-Saber}{2006a}]%
        {Saber2006}
\bibfield{author}{\bibinfo{person}{R. Olfati-Saber}.}
  \bibinfo{year}{2006}\natexlab{a}.
\newblock \showarticletitle{Flocking for multi-agent dynamic systems:
  algorithms and theory}.
\newblock \bibinfo{journal}{{\it IEEE Trans. Automat. Control}}
  \bibinfo{volume}{51}, \bibinfo{number}{3} (\bibinfo{year}{2006}),
  \bibinfo{pages}{401--420}.
\newblock
\showDOI{%
\url{https://doi.org/10.1109/tac.2005.864190}}


\bibitem[\protect\citeauthoryear{Olfati-Saber}{Olfati-Saber}{2006b}]%
        {Saber2004}
\bibfield{author}{\bibinfo{person}{R. Olfati-Saber}.}
  \bibinfo{year}{2006}\natexlab{b}.
\newblock \showarticletitle{Flocking for multi-agent dynamic systems:
  Algorithms and theory}.
\newblock \bibinfo{journal}{{\it IEEE Trans. Automat. Control}}
  \bibinfo{volume}{51}, \bibinfo{number}{3} (\bibinfo{year}{2006}),
  \bibinfo{pages}{401--420}.
\newblock


\bibitem[\protect\citeauthoryear{Olfati-Saber and Murray}{Olfati-Saber and
  Murray}{2002}]%
        {olfati2002distributed}
\bibfield{author}{\bibinfo{person}{R. Olfati-Saber} {and}
  \bibinfo{person}{R.~M. Murray}.} \bibinfo{year}{2002}\natexlab{}.
\newblock \showarticletitle{Distributed cooperative control of multiple vehicle
  formations using structural potential functions}.
\newblock \bibinfo{journal}{{\em IFAC Proceedings Volumes\/}}
  \bibinfo{volume}{35}, \bibinfo{number}{1} (\bibinfo{year}{2002}),
  \bibinfo{pages}{495--500}.
\newblock


\bibitem[\protect\citeauthoryear{Qin, Li, and Shen}{Qin et~al\mbox{.}}{2018}]%
        {VINS}
\bibfield{author}{\bibinfo{person}{T. Qin}, \bibinfo{person}{P. Li}, {and}
  \bibinfo{person}{S. Shen}.} \bibinfo{year}{2018}\natexlab{}.
\newblock \showarticletitle{Vins-mono: a robust and versatile monocular
  visual-inertial state estimator}.
\newblock \bibinfo{journal}{{\em IEEE Transactions on Robotics\/}}
  \bibinfo{volume}{34}, \bibinfo{number}{4} (\bibinfo{year}{2018}),
  \bibinfo{pages}{1004--1020}.
\newblock


\bibitem[\protect\citeauthoryear{Ren and Beard}{Ren and Beard}{2005}]%
        {ren2005consensus}
\bibfield{author}{\bibinfo{person}{W. Ren} {and} \bibinfo{person}{R.~W.
  Beard}.} \bibinfo{year}{2005}\natexlab{}.
\newblock \showarticletitle{Consensus seeking in multiagent systems under
  dynamically changing interaction topologies}.
\newblock \bibinfo{journal}{{\em IEEE Transactions on automatic control\/}}
  \bibinfo{volume}{50}, \bibinfo{number}{5} (\bibinfo{year}{2005}),
  \bibinfo{pages}{655--661}.
\newblock


\bibitem[\protect\citeauthoryear{Ren and Sorensen}{Ren and Sorensen}{2008}]%
        {Virtual2008}
\bibfield{author}{\bibinfo{person}{W. Ren} {and} \bibinfo{person}{N.
  Sorensen}.} \bibinfo{year}{2008}\natexlab{}.
\newblock \showarticletitle{Distributed coordination architecture for
  multi-robot formation control}.
\newblock \bibinfo{journal}{{\em Robotics and Autonomous Systems\/}}
  \bibinfo{volume}{56}, \bibinfo{number}{4} (\bibinfo{year}{2008}),
  \bibinfo{pages}{324--333}.
\newblock


\bibitem[\protect\citeauthoryear{Reynolds}{Reynolds}{1987}]%
        {Reynolds1987}
\bibfield{author}{\bibinfo{person}{C.~W. Reynolds}.}
  \bibinfo{year}{1987}\natexlab{}.
\newblock \showarticletitle{Flocks, herds and schools: a distributed behavioral
  model}.
\newblock  \bibinfo{volume}{21}, \bibinfo{number}{4} (\bibinfo{year}{1987}).
\newblock


\bibitem[\protect\citeauthoryear{Romero-Ramirez, Mu{\~n}oz-Salinas, and
  Medina-Carnicer}{Romero-Ramirez et~al\mbox{.}}{2018}]%
        {Aruco}
\bibfield{author}{\bibinfo{person}{F.~J. Romero-Ramirez}, \bibinfo{person}{R.
  Mu{\~n}oz-Salinas}, {and} \bibinfo{person}{R. Medina-Carnicer}.}
  \bibinfo{year}{2018}\natexlab{}.
\newblock \showarticletitle{Speeded up detection of squared fiducial markers}.
\newblock \bibinfo{journal}{{\em Image and vision Computing\/}}
  \bibinfo{volume}{76} (\bibinfo{year}{2018}), \bibinfo{pages}{38--47}.
\newblock


\bibitem[\protect\citeauthoryear{Saska, Vakula, and P{\v{r}}eu{\'c}il}{Saska
  et~al\mbox{.}}{2014}]%
        {Martin2014}
\bibfield{author}{\bibinfo{person}{M. Saska}, \bibinfo{person}{J. Vakula},
  {and} \bibinfo{person}{L. P{\v{r}}eu{\'c}il}.}
  \bibinfo{year}{2014}\natexlab{}.
\newblock \showarticletitle{Swarms of micro aerial vehicles stabilized under a
  visual relative localization}.
\newblock \bibinfo{journal}{{\em IEEE International Conference on Robotics and
  Automation (ICRA)\/}} (\bibinfo{year}{2014}), \bibinfo{pages}{3570--3575}.
\newblock


\bibitem[\protect\citeauthoryear{Tanner, Jadbabaie, and Pappas}{Tanner
  et~al\mbox{.}}{2003a}]%
        {FixedTopology}
\bibfield{author}{\bibinfo{person}{H.~G. Tanner}, \bibinfo{person}{A.
  Jadbabaie}, {and} \bibinfo{person}{G.~J. Pappas}.}
  \bibinfo{year}{2003}\natexlab{a}.
\newblock \showarticletitle{Stable flocking of mobile agents, {Part I}: Fixed
  topology}. In \bibinfo{booktitle}{{\em IEEE International Conference on
  Decision and Control}}, Vol.~\bibinfo{volume}{2}. IEEE,
  \bibinfo{pages}{2010--2015}.
\newblock


\bibitem[\protect\citeauthoryear{Tanner, Jadbabaie, and Pappas}{Tanner
  et~al\mbox{.}}{2003b}]%
        {DynamicTopology}
\bibfield{author}{\bibinfo{person}{H.~G. Tanner}, \bibinfo{person}{A.
  Jadbabaie}, {and} \bibinfo{person}{G.~J. Pappas}.}
  \bibinfo{year}{2003}\natexlab{b}.
\newblock \showarticletitle{Stable flocking of mobile agents {Part II}: Dynamic
  topology}. In \bibinfo{booktitle}{{\em IEEE International Conference on
  Decision and Control}}, Vol.~\bibinfo{volume}{2}. IEEE,
  \bibinfo{pages}{2016--2021}.
\newblock


\bibitem[\protect\citeauthoryear{Tanner, Jadbabaie, and Pappas}{Tanner
  et~al\mbox{.}}{2007}]%
        {tanner2007flocking}
\bibfield{author}{\bibinfo{person}{H.~G. Tanner}, \bibinfo{person}{A.
  Jadbabaie}, {and} \bibinfo{person}{G.~J. Pappas}.}
  \bibinfo{year}{2007}\natexlab{}.
\newblock \showarticletitle{Flocking in fixed and switching networks}.
\newblock \bibinfo{journal}{{\em IEEE Transactions on Automatic control\/}}
  \bibinfo{volume}{52}, \bibinfo{number}{5} (\bibinfo{year}{2007}),
  \bibinfo{pages}{863--868}.
\newblock


\bibitem[\protect\citeauthoryear{V{\'a}s{\'a}rhelyi, Vir{\'a}gh, Somorjai,
  Nepusz, Eiben, and Vicsek}{V{\'a}s{\'a}rhelyi et~al\mbox{.}}{2018}]%
        {Vicsek2018}
\bibfield{author}{\bibinfo{person}{G. V{\'a}s{\'a}rhelyi}, \bibinfo{person}{C.
  Vir{\'a}gh}, \bibinfo{person}{G. Somorjai}, \bibinfo{person}{T. Nepusz},
  \bibinfo{person}{A.~E. Eiben}, {and} \bibinfo{person}{T. Vicsek}.}
  \bibinfo{year}{2018}\natexlab{}.
\newblock \showarticletitle{Optimized flocking of autonomous drones in confined
  environments}.
\newblock \bibinfo{journal}{{\em Science Robotics\/}} \bibinfo{volume}{3},
  \bibinfo{number}{20} (\bibinfo{year}{2018}), \bibinfo{pages}{eaat3536}.
\newblock


\bibitem[\protect\citeauthoryear{Vicsek, Czir{\'o}k, Ben-Jacob, Cohen, and
  Shochet}{Vicsek et~al\mbox{.}}{1995}]%
        {Vicsek1995}
\bibfield{author}{\bibinfo{person}{T. Vicsek}, \bibinfo{person}{A. Czir{\'o}k},
  \bibinfo{person}{E. Ben-Jacob}, \bibinfo{person}{I. Cohen}, {and}
  \bibinfo{person}{Shochet}.} \bibinfo{year}{1995}\natexlab{}.
\newblock \showarticletitle{Novel type of phase transition in a system of
  self-driven particles}.
\newblock \bibinfo{journal}{{\em Physical review letters\/}}
  \bibinfo{volume}{75}, \bibinfo{number}{6} (\bibinfo{year}{1995}),
  \bibinfo{pages}{1226}.
\newblock


\bibitem[\protect\citeauthoryear{Virágh, Vásárhelyi, Tarcai, Szörényi, G.,
  Nepusz, and Vicsek}{Virágh et~al\mbox{.}}{2014}]%
        {Vicsek2014}
\bibfield{author}{\bibinfo{person}{C. Virágh}, \bibinfo{person}{G.
  Vásárhelyi}, \bibinfo{person}{N. Tarcai}, \bibinfo{person}{T. Szörényi},
  \bibinfo{person}{Somorjai G.}, \bibinfo{person}{T. Nepusz}, {and}
  \bibinfo{person}{T. Vicsek}.} \bibinfo{year}{2014}\natexlab{}.
\newblock \showarticletitle{Flocking algorithm for autonomous flying robots}.
\newblock \bibinfo{journal}{{\em Bioinspiration and Biomimetics\/}}
  \bibinfo{volume}{9}, \bibinfo{number}{2} (\bibinfo{year}{2014}),
  \bibinfo{pages}{025012}.
\newblock
\showDOI{%
\url{https://doi.org/10.1088/1748-3182/9/2/025012}}


\bibitem[\protect\citeauthoryear{Young, Scardovi, Cavagna, Giardina, and
  Leonard}{Young et~al\mbox{.}}{2013}]%
        {lowcost}
\bibfield{author}{\bibinfo{person}{G.~F. Young}, \bibinfo{person}{L. Scardovi},
  \bibinfo{person}{A. Cavagna}, \bibinfo{person}{I. Giardina}, {and}
  \bibinfo{person}{N.~E. Leonard}.} \bibinfo{year}{2013}\natexlab{}.
\newblock \showarticletitle{Starling flock networks manage uncertainty in
  consensus at low cost}.
\newblock \bibinfo{journal}{{\em PLoS Computational Biology\/}}
  \bibinfo{volume}{9}, \bibinfo{number}{1} (\bibinfo{year}{2013}).
\newblock
\showDOI{%
\url{https://doi.org/10.1371/journal.pcbi.1002894}}


\bibitem[\protect\citeauthoryear{Özaslan, Shen, Mulgaonkar, Michael, and
  Kumar}{Özaslan et~al\mbox{.}}{2015}]%
        {inspection}
\bibfield{author}{\bibinfo{person}{T. Özaslan}, \bibinfo{person}{S. Shen},
  \bibinfo{person}{Y. Mulgaonkar}, \bibinfo{person}{N. Michael}, {and}
  \bibinfo{person}{V. Kumar}.} \bibinfo{year}{2015}\natexlab{}.
\newblock \showarticletitle{Inspection of penstocks and featureless tunnel-like
  environments using micro UAVs}.
\newblock \bibinfo{journal}{{\em Springer Tracts in Advanced Robotics Field and
  Service Robotics\/}} (\bibinfo{year}{2015}), \bibinfo{pages}{123–136}.
\newblock
\showDOI{%
\url{https://doi.org/10.1007/978-3-319-07488-7_9}}


\end{thebibliography}
